# Transverse spin effects in hadron-pair production from semi-inclusive deep inelastic scattering

The COMPASS Collaboration

**Abstract**

First measurements of azimuthal asymmetries in hadron-pair production in deep-inelastic scattering of muons on transversely polarised $^6$LiD (deuteron) and $NH_3$ (proton) targets are presented. The data were taken in the years 2002–2004 and 2007 with the COMPASS spectrometer using a muon beam of 160 GeV/$c$ at the CERN SPS. The asymmetries provide access to the transversity distribution functions, without involving the Collins effect as in single hadron production. The sizeable asymmetries measured on the $NH_3$ target indicate non-vanishing $u$-quark transversity and two-hadron interference fragmentation functions. The small asymmetries measured on the $^6$LiD target can be interpreted as indication for a cancellation of $u$- and $d$-quark transversities.

## The COMPASS Collaboration

C. Adolph[8], M.G. Alekseev[28], V.Yu. Alexakhin[7], Yu. Alexandrov[15,*], G.D. Alexeev[7], A. Amoroso[27], A.A. Antonov[7], A. Austregesilo[10,17], B. Badełek[30], F. Balestra[27], J. Barth[4], G. Baum[1], Y. Bedfer[22], J. Bernhard[13], R. Bertini[27], M. Bettinelli[16], K. Bicker[10,17], J. Bieling[4], R. Birsa[24], J. Bisplinghoff[3], P. Bordalo[12,a], F. Bradamante[25], C. Braun[8], A. Bravar[24], A. Bressan[25], E. Burtin[22], D. Chaberny[13], M. Chiosso[27], S.U. Chung[17], A. Cicuttin[26], M.L. Crespo[26], S. Dalla Torre[24], S. Das[6], S.S. Dasgupta[6], O.Yu. Denisov[10,28], L. Dhara[6], S.V. Donskov[21], N. Doshita[2,32], V. Duic[25], W. Dünnweber[16], M. Dziewiecki[31], A. Efremov[7], C. Elia[25], P.D. Eversheim[3], W. Eyrich[8], M. Faessler[16], A. Ferrero[22], A. Filin[21], M. Finger[19], M. Finger jr.[7], H. Fischer[9], C. Franco[12], N. du Fresne von Hohenesche[13,10], J.M. Friedrich[17], R. Garfagnini[27], F. Gautheron[2], O.P. Gavrichtchouk[7], R. Gazda[30], S. Gerassimov[15,17], R. Geyer[16], M. Giorgi[25], I. Gnesi[27], B. Gobbo[24], S. Goertz[2,4], S. Grabmüller[17], A. Grasso[27], B. Grube[17], R. Gushterski[7], A. Guskov[7], T. Guthörl[9], F. Haas[17], D. von Harrach[13], F.H. Heinsius[9], F. Herrmann[9], C. Heß[2], F. Hinterberger[3], N. Horikawa[18,b], Ch. Höppner[17], N. d'Hose[22], S. Huber[17] S. Ishimoto[18,c], O. Ivanov[7], Yu. Ivanshin[7], T. Iwata[32], R. Jahn[3], P. Jasinski[13], R. Joosten[3], E. Kabuß[13], D. Kang[13], B. Ketzer[17], G.V. Khaustov[21], Yu.A. Khokhlov[21], Yu. Kisselev[2], F. Klein[4], K. Klimaszewski[30], S. Koblitz[13], J.H. Koivuniemi[2], V.N. Kolosov[21], K. Kondo[2,32], K. Königsmann[9], I. Konorov[15,17], V.F. Konstantinov[21], A. Korzenev[22,d], A.M. Kotzinian[27], O. Kouznetsov[7,22], M. Krämer[17], Z.V. Kroumchtein[7], F. Kunne[22], K. Kurek[30], L. Lauser[9], A.A. Lednev[21], A. Lehmann[8], S. Levorato[25], J. Lichtenstadt[23], A. Maggiora[28], A. Magnon[22], N. Makke[22,25], G.K. Mallot[10], A. Mann[17], C. Marchand[22], A. Martin[25], J. Marzec[31], F. Massmann[3], T. Matsuda[14], W. Meyer[2], T. Michigami[32], A. Mielech[25,30], Yu.V. Mikhailov[21], M.A. Moinester[23], A. Morreale[22], A. Mutter[9,13], A. Nagaytsev[7], T. Nagel[17], T. Negrini[9], F. Nerling[9], S. Neubert[17], D. Neyret[22], V.I. Nikolaenko[21], W.D. Nowak[9], A.S. Nunes[12], A.G. Olshevsky[7], M. Ostrick[13], A. Padee[31], R. Panknin[4], D. Panzieri[29], B. Parsamyan[27], S. Paul[17], E. Perevalova[7], G. Pesaro[25], D.V. Peshekhonov[7], G. Piragino[27], S. Platchkov[22], J. Pochodzalla[13], J. Polak[11,25], V.A. Polyakov[21], G. Pontecorvo[7], J. Pretz[4], M. Quaresma[12], C. Quintans[12], J.-F. Rajotte[16], S. Ramos[12,a], V. Rapatsky[7], G. Reicherz[2], A. Richter[8], E. Rocco[10], E. Rondio[30], N.S. Rossiyskaya[7], D.I. Ryabchikov[21], V.D. Samoylenko[21], A. Sandacz[30], M.G. Sapozhnikov[7], S. Sarkar[6], I.A. Savin[7], G. Sbrizzai[25], P. Schiavon[25], C. Schill[9], T. Schlüter[16], K. Schmidt[9], L. Schmitt[17,e], K. Schönning[10], S. Schopferer[9], M. Schott[10], O.Yu. Shevchenko[7], L. Silva[12], L. Sinha[6], A.N. Sissakian[7,*], M. Slunecka[7], G.I. Smirnov[7], S. Sosio[27], F. Sozzi[24], A. Srnka[5], M. Stolarski[12], M. Sulc[11], R. Sulej[30], P. Sznajder[30], S. Takekawa[25], J. Ter Wolbeek[9], S. Tessaro[24], F. Tessarotto[24], L.G. Tkatchev[7], S. Uhl[17], I. Uman[16], M. Vandenbroucke[22], M. Virius[20], N.V. Vlassov[7], A. Vossen[9], L. Wang[2], R. Windmolders[4], W. Wiślicki[30], H. Wollny[9,22], K. Zaremba[31], M. Zavertyaev[15], E. Zemlyanichkina[7], M. Ziembicki[31], N. Zhuravlev[7] and A. Zvyagin[16]

[1] Universität Bielefeld, Fakultät für Physik, 33501 Bielefeld, Germany[f]
[2] Universität Bochum, Institut für Experimentalphysik, 44780 Bochum, Germany[f]
[3] Universität Bonn, Helmholtz-Institut für Strahlen- und Kernphysik, 53115 Bonn, Germany[f]
[4] Universität Bonn, Physikalisches Institut, 53115 Bonn, Germany[f]
[5] Institute of Scientific Instruments, AS CR, 61264 Brno, Czech Republic[g]
[6] Matrivani Institute of Experimental Research & Education, Calcutta-700 030, India[h]
[7] Joint Institute for Nuclear Research, 141980 Dubna, Moscow region, Russia[i]
[8] Universität Erlangen–Nürnberg, Physikalisches Institut, 91054 Erlangen, Germany[f]
[9] Universität Freiburg, Physikalisches Institut, 79104 Freiburg, Germany[f]
[10] CERN, 1211 Geneva 23, Switzerland
[11] Technical University in Liberec, 46117 Liberec, Czech Republic[g]
[12] LIP, 1000-149 Lisbon, Portugal[j]
[13] Universität Mainz, Institut für Kernphysik, 55099 Mainz, Germany[f]
[14] University of Miyazaki, Miyazaki 889-2192, Japan[k]

[15] Lebedev Physical Institute, 119991 Moscow, Russia
[16] Ludwig-Maximilians-Universität München, Department für Physik, 80799 Munich, Germany[f,l)]
[17] Technische Universität München, Physik Department, 85748 Garching, Germany[f,l)]
[18] Nagoya University, 464 Nagoya, Japan[k]
[19] Charles University in Prague, Faculty of Mathematics and Physics, 18000 Prague, Czech Republic[g]
[20] Czech Technical University in Prague, 16636 Prague, Czech Republic[g]
[21] State Research Center of the Russian Federation, Institute for High Energy Physics, 142281 Protvino, Russia
[22] CEA IRFU/SPhN Saclay, 91191 Gif-sur-Yvette, France
[23] Tel Aviv University, School of Physics and Astronomy, 69978 Tel Aviv, Israel[m]
[24] Trieste Section of INFN, 34127 Trieste, Italy
[25] University of Trieste, Department of Physics and Trieste Section of INFN, 34127 Trieste, Italy
[26] Abdus Salam ICTP and Trieste Section of INFN, 34127 Trieste, Italy
[27] University of Turin, Department of Physics and Torino Section of INFN, 10125 Turin, Italy
[28] Torino Section of INFN, 10125 Turin, Italy
[29] University of Eastern Piedmont, 1500 Alessandria, and Torino Section of INFN, 10125 Turin, Italy
[30] National Center for Nuclear Research and University of Warsaw, 00-681 Warsaw, Poland[n]
[31] Warsaw University of Technology, Institute of Radioelectronics, 00-665 Warsaw, Poland[n]
[32] Yamagata University, Yamagata, 992-8510 Japan[k]
[a] Also at IST, Universidade Técnica de Lisboa, Lisbon, Portugal
[b] Also at Chubu University, Kasugai, Aichi, 487-8501 Japan[k]
[c] Also at KEK, 1-1 Oho, Tsukuba, Ibaraki, 305-0801 Japan
[d] On leave of absence from JINR Dubna
[e] Also at GSI mbH, Planckstr. 1, D-64291 Darmstadt, Germany
[f] Supported by the German Bundesministerium für Bildung und Forschung
[g] Suppported by Czech Republic MEYS grants ME492 and LA242
[h] Supported by SAIL (CSR), Govt. of India
[i] Supported by CERN-RFBR grants 08-02-91009
[j] Supported by the Portuguese FCT - Fundação para a Ciência e Tecnologia, COMPETE and QREN, grants CERN/FP/83542/2008, CERN/FP/109323/2009 and CERN/FP/116376/2010
[k] Supported by the MEXT and the JSPS under the Grants No.18002006, No.20540299 and No.18540281; Daiko Foundation and Yamada Foundation
[l] Supported by the DFG cluster of excellence 'Origin and Structure of the Universe' (www.universe-cluster.de)
[m] Supported by the Israel Science Foundation, founded by the Israel Academy of Sciences and Humanities
[n] Supported by Ministry of Science and Higher Education grant 41/N-CERN/2007/0
[*] Deceased

## 1 Introduction

The quark content of the nucleon at twist-two level can be characterised by three independent distribution functions (DFs) for each quark flavour[1]. If the quarks are collinear with the parent nucleon, i.e. quarks have no intrinsic transverse momentum $\boldsymbol{k_T}$, or after integration over $\boldsymbol{k_T}$, these three distributions exhaust the information on the partonic structure of the nucleon [2–5]. Two of these functions have been measured with good accuracy. First, the unpolarised distribution $f_1^q(x)$ describes the number density of quarks with flavour $q$ carrying a fraction $x$ of the longitudinal momentum of the fast moving parent nucleon. Secondly, the helicity distribution $g_1^q(x)$ describes the difference between the number densities of quarks with helicity parallel and antiparallel to the spin of the nucleon that is longitudinally polarised with respect to the nucleon momentum. However, up to ten years ago nothing was known about the third function, the transverse spin distribution $h_1^q(x)$, often referred to as transversity, which describes the difference between the number densities of quarks with their spins oriented parallel and antiparallel to the spin of a transversely polarised nucleon. This distribution is difficult to measure, since it is related to soft processes correlating quarks with opposite chirality, making it a chiral-odd function [1]. As a result, transversity can only be accessed through observables in which it appears coupled with a second chiral-odd object in order to conserve chirality, i.e. not in inclusive deep inelastic scattering (DIS) at leading twist. For a recent review see Ref. [6].

The only results on transversity so far come from semi-inclusive deep inelastic scattering (SIDIS) reactions. Here, the chiral-odd partners of the transversity distribution function are fragmentation functions (FFs), which describe the spin-dependent hadronisation of a transversely polarised quark $q$ into hadrons. Already in 1993 it was suggested by Collins [7] that transversity could be accessed in SIDIS due to a hadronisation mechanism, in which the transverse spin of the struck quark is correlated to the transverse momentum $\boldsymbol{p_T}$ of the produced unpolarised hadron with respect to the quark momentum. The resulting azimuthal asymmetry in the distribution of the final state hadrons is then proportional to the convolution of this Collins FF $H_1^h(z, p_T)$ and the transversity DF $h_1^q(x, k_T)$. Here, convolution integrals of quark and hadron transverse momenta have to be taken into account [3, 4, 8].

Such asymmetries were measured by HERMES using a transversely polarised pure hydrogen target [9] and by COMPASS using transversely polarised $^6$LiD (deuteron) [10] and $NH_3$ (proton) targets [11]. No significant signals were found on the $^6$LiD target. Sizeable Collins asymmetries were observed by HERMES, and by COMPASS using the polarised $NH_3$ target, which implies both transversity DF and Collins FF to be different from zero. The latter was also independently measured at the KEK $e^+e^-$ collider by the BELLE experiment and established to be sizeable [12, 13]. All these results were well described by a global fit [14, 15], which allowed for a first extraction of the $u$- and $d$-quark transversity DFs. These results contain convolutions of transverse-momentum-dependent functions, where the assumed shape of the transverse momentum distributions leads to a model dependence of the extracted results. Moreover, BELLE results as compared to COMPASS and HERMES results were obtained at a very different scale, $Q^2 \approx 100\,(\mathrm{GeV}/c)^2$ compared to $\langle Q^2 \rangle \approx 3.2\,(\mathrm{GeV}/c)^2$ and $\langle Q^2 \rangle \approx 2.4\,(\mathrm{GeV}/c)^2$, respectively. The neccessary $Q^2$ evolution is quite complex for transverse-momentum-dependent functions [8, 16], requiring the use of Collins-Soper factorisation [17, 18] and thus again the knowledge of the transverse momentum distributions. In the global analysis of Refs. [14, 15], a gaussian ansatz for the transverse momentum distributions was used but only collinear evolution was taken into account, leading to a possible overestimation of $h_1$ [8].

As an alternative to study the Collins fragmentation mechanism in one-hadron SIDIS it was proposed to measure two-hadron SIDIS, $lp^\uparrow \to l'h^+h^-X$ with both hadrons produced in the current fragmentation region [19–22]. In this reaction appears a new chiral odd fragmentation function, the Interference Fragmentation Function (IFF) $H_1^\sphericalangle$, which describes the fragmentation of a transversely polarised quark into a pair of unpolarised hadrons. The transverse polarisation of the fragmenting quark is correlated with the relative momentum of the two hadrons, which gives rise to an azimuthal asymmetry with respect to the virtual-photon direction and the lepton scattering plane. The additional degrees of freedom allow for an integration over the transverse momenta of the final state hadrons, leaving only the relative momentum of the two hadrons. This avoids the complexity of transverse-momentum-dependent convolution integrals and the analysis can be performed using collinear factorisation [8, 16]. Here, the evolution equations are known at next-to-leading order [23], so that results from $e^+e^-$ scattering and SIDIS can be correctly connected, making it the theoretically cleanest way to extract transversity using presently existing facilities [8]. The properties of interference fragmentation functions are described in detail in Refs. [19–22, 24–27].
First evidence for azimuthal asymmetries in leptoproduction of $\pi^+\pi^-$ pairs was published by HERMES, using a transversely polarised hydrogen target [28]. The interference fragmentation functions were measured in $e^+e^-$ reactions by BELLE [29]. These measurements indicate a sizeable $u$-quark transversity distribution and non-vanishing interference fragmentation functions [30]. In this Letter, the first measurement of two-hadron azimuthal asymmetries using a transversely polarised $^6$LiD (deuteron) target is presented as well as results from a $NH_3$ (proton) target. Due to the large acceptance of the COMPASS spectrometer and the large beam momentum of 160 GeV/$c$, results with high statistics were obtained covering a large kinematic range in $x$ and $M_{hh}$, the invariant mass of the hadron pair.

## 2 Theoretical Framework

At leading twist and after integration over the hadron transverse momenta, the cross section of semi-inclusive two-hadron leptoproduction on a transversely polarised target is given as a sum of a spin independent and a spin dependent part [25, 26]:

$$\frac{d^7\sigma_{UU}}{d\cos\theta\, dM_{hh}^2\, d\phi_R\, dz\, dx\, dy\, d\phi_S} = \frac{\alpha^2}{2\pi Q^2 y}\left(1-y+\frac{y^2}{2}\right) \times \sum_q e_q^2 f_1^q(x)\, D_{1,q}\big(z, M_{hh}^2, \cos\theta\big), \tag{1}$$

$$\frac{d^7\sigma_{UT}}{d\cos\theta\, dM_{hh}^2\, d\phi_R\, dz\, dx\, dy\, d\phi_S} = \frac{\alpha^2}{2\pi Q^2 y} S_\perp (1-y) \times \sum_q e_q^2 \frac{|\boldsymbol{p_1}-\boldsymbol{p_2}|}{2M_{hh}} \sin\theta\, \sin\phi_{RS}\, h_1^q(x)\, H_{1,q}^{\sphericalangle}\big(z, M_{hh}^2, \cos\theta\big). \tag{2}$$

Here, the sums run over all quark flavours $q$, and $\boldsymbol{p_1}$ and $\boldsymbol{p_2}$ denote the momenta of the two hadrons of the pair. The first index ($U$) indicates an unpolarised beam and the second, ($U$) or ($T$), an unpolarised and transversely polarised target, respectively. Note that the contribution from a longitudinally polarised beam and a transversely polarised target, $\sigma_{LT}$, is suppressed by $1/Q$ [31]. The fine-structure constant is denoted by $\alpha$, $y$ is the fraction of the muon's energy lost in the nucleon rest frame, $D_{1,q}(z, M_{hh}^2, \cos\theta)$ is the unpolarised two-hadron fragmentation function for a quark of flavour $q$ and $z_1$, $z_2$ are the fractions of the virtual-photon energy carried by these two hadrons, with $z = z_1 + z_2$. The symbol $S_\perp$ denotes the component of the target spin $\boldsymbol{S}$ perpendicular to the virtual-photon direction, and $\theta$ is the polar angle of one of the hadrons – commonly the positive one – in the two-hadron rest frame with respect to the two-hadron boost axis. The azimuthal angle $\phi_{RS}$ is defined according to Fig. 1 as

$$\phi_{RS} = \phi_R - \phi_{S'} = \phi_R + \phi_S - \pi\,, \tag{3}$$

where $\phi_S$ is the azimuthal angle of the initial nucleon spin and $\phi_{S'}$ is the azimuthal angle of the spin of the fragmenting quark, with $\phi_{S'} = \pi - \phi_S$. The azimuthal angle $\phi_R$ is defined by

$$\phi_R = \frac{(\boldsymbol{q}\times\boldsymbol{l})\cdot\boldsymbol{R}}{|(\boldsymbol{q}\times\boldsymbol{l})\cdot\boldsymbol{R}|} \arccos\left(\frac{(\boldsymbol{q}\times\boldsymbol{l})\cdot(\boldsymbol{q}\times\boldsymbol{R})}{|\boldsymbol{q}\times\boldsymbol{l}||\boldsymbol{q}\times\boldsymbol{R}|}\right), \tag{4}$$

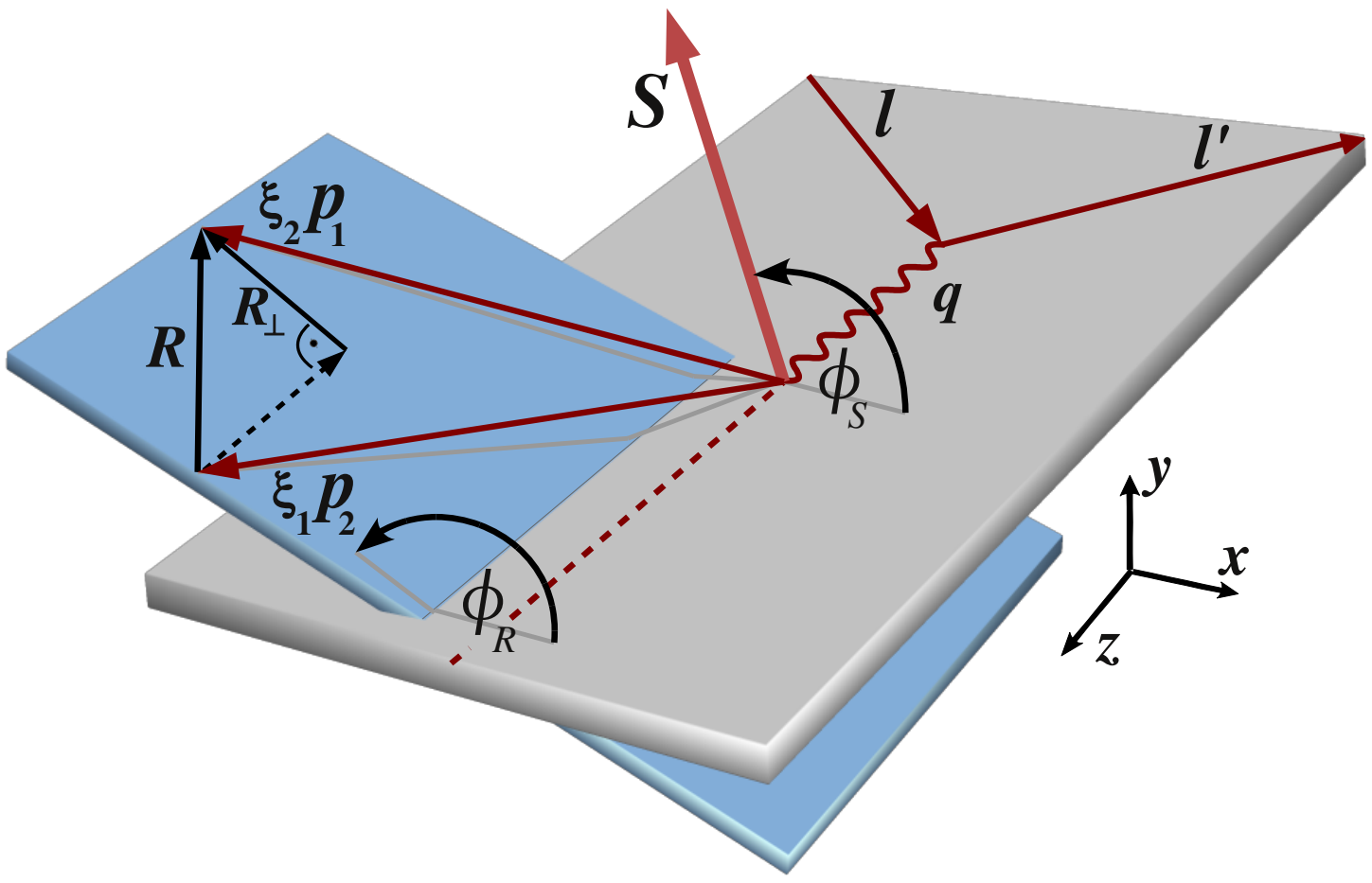

Fig. 1: Definition of the azimuthal angles $\phi_R$ and $\phi_S$ for two-hadron production in deep inelastic scattering, where $\boldsymbol{l}$, $\boldsymbol{l'}$, $\boldsymbol{q}$ and $\boldsymbol{p_i}$ are the 3-momenta of beam, scattered muon, virtual photon and hadrons. Note that the azimuthal plane is defined by the directions of the relative hadron momentum and the virtual photon.

where $\boldsymbol{l}$ is the incoming lepton momentum, $\boldsymbol{q}$ the virtual-photon momentum and $\boldsymbol{R}$ the relative hadron momentum [20, 32] given by

$$\boldsymbol{R} = \frac{z_2\boldsymbol{p_1} - z_1\boldsymbol{p_2}}{z_1 + z_2} =: \xi_2\boldsymbol{p_1} - \xi_1\boldsymbol{p_2}\,. \tag{5}$$

The number $N_{h^+h^-}$ of pairs of oppositely charged hadrons produced on a transversely polarised target can be written as

$$N_{h^+h^-}(x,y,z,M_{hh}^2,\cos\theta,\phi_{RS}) \propto \sigma_{UU}\left(1 \pm f(x,y)\,P_T\,D_{nn}(y)\,A_{UT}^{\sin\phi_{RS}}\sin\theta\sin\phi_{RS}\right), \tag{6}$$

omitting luminosity and detector acceptance. Here, $P_T$ is the magnitude of the transverse target polarisation and $D_{nn}(y) = \frac{1-y}{1-y+y^2/2}$ the transverse-spin-transfer coefficient, while $f(x,y)$ is the target dilution factor calculated for semi-inclusive reactions that depends on kinematics. It is given by the number-weighted ratio of the total cross section for scattering on protons or deuterons to that for scattering on all nuclei in the target. It increases at large $x$ due to the reduced cross section for heavy targets. The dependence of the dilution factor on the hadron transverse momenta appears to be weak in the $\nu$ range of the COMPASS experiment. Dilution due to radiative events on protons or deuterons is taken into account by the ratio of the one-photon exchange cross section to the total cross section. For $^6$LiD, $f$ includes a correction for the relative polarisation of deuterons bound in $^6$Li with respect to free deuterons. For $^{14}$NH$_3$, $f$ contains a correction for the polarisation of the small $^{15}$N admixture.

The asymmetry

$$A_{UT}^{\sin\phi_{RS}} = \frac{|\boldsymbol{p_1} - \boldsymbol{p_2}|}{2M_{hh}} \frac{\sum_q e_q^2 \cdot h_1^q(x) \cdot H_{1,q}^{\sphericalangle}(z, M_{hh}^2, \cos\theta)}{\sum_q e_q^2 \cdot f_1^q(x) \cdot D_{1,q}(z, M_{hh}^2, \cos\theta)} \tag{7}$$

is proportional to the product of the transversity distribution function $h_1^q(x)$ and the polarised two-hadron interference fragmentation function $H_{1,q}^{\sphericalangle}(z, M_{hh}^2, \cos\theta)$, summed over the quark flavours $q$ with charge $e_q$.

It is convenient [25] to expand both the polarised and unpolarised two-hadron fragmentation functions in terms of Legendre polynomials in $\cos\theta$. For the invariant mass range typically covered by SIDIS experiments, $M_{hh} < 1.5\,\mathrm{GeV}/c^2$, to a good approximation only relative s and p partial waves of the two-hadron system contribute to the cross section, yielding [27]

$$\begin{aligned} D_1(z, M_{hh}^2, \cos\theta) \;\simeq\; & D_{1,oo}(z, M_{hh}^2) + \cos\theta\, D_{1,ol}(z, M_{hh}^2) \\ & + \frac{1}{4}(3\cos^2\theta - 1)\, D_{1,ll}(z, M_{hh}^2) \end{aligned} \tag{8}$$

and

$$H_1^{\sphericalangle}(z, M_{hh}^2, \cos\theta) \simeq H_{1,ot}^{\sphericalangle}(z, M_{hh}^2) + \cos\theta H_{1,lt}^{\sphericalangle}(z, M_{hh}^2), \tag{9}$$

respectively. The term $\cos\theta D_{1,ol}(z, M_{hh}^2)$ describes the interference between an unpolarised hadron pair (denoted $o$) in s-wave and a longitudinally polarised pair (denoted $l$) in p-wave, while the term $H_{1,ot}^{\sphericalangle}(z, M_{hh}^2)$ arises from the interference between an unpolarised hadron pair in s-wave and a transversely polarised pair (denoted $t$) in p-wave. The term $\cos\theta H_{1,lt}^{\sphericalangle}(z, M_{hh}^2)$ indicates interference between longitudinally and transversely polarised pairs in a relative p-wave, while the term $(3\cos^2\theta - 1)D_{1,ll}(z, M_{hh}^2)$ indicates interference between longitudinally polarised pairs in a relative p-wave. The term $D_{1,oo}(z, M_{hh}^2)$ represents an unpolarised state of the hadron pair and can have contributions from s- and p-waves but not from the interference between both.

## 3 Experimental Data

The analysis presented in this Letter is performed using data taken in the years 2002–2004 and 2007 with the COMPASS spectrometer [33] by scattering positive muons of $160\,\mathrm{GeV}/c$ from the CERN SPS off transversely polarised solid state $^6$LiD and $NH_3$ targets, respectively. The beam muons originating from $\pi^+$ and $K^+$ decays are naturally polarised with an average longitudinal polarisation of about 0.8 with a relative uncertainty of 5 %. For $^6$LiD, the average dilution factor calculated for semi-inclusive reactions is $\langle f \rangle \sim 0.38$ and the average polarisation is $\langle P_T \rangle \sim 0.47$, while for $NH_3$ it is $\langle f \rangle \sim 0.15$ and $\langle P_T \rangle \sim 0.83$, respectively. The target consists of cylindrical cells in a row, which can be independently polarised. In 2002–2004, two cells were used, each 60 cm long and 3 cm in diameter. The direction of polarisation in the downstream cell was chosen oppositely to the one in the upstream cell. In 2007, the target consisted of three cylindrical cells, with 4 cm diameter. The middle cell was 60 cm long and the two outer cells 30 cm each. The direction of polarisation in the middle cell was opposite to the one in the outer cells. For the analysis the central cell is divided into two parts, providing four data samples with two different orientations of polarisation. Both target configurations allow for a simultaneous measurement of azimuthal asymmetries for both target spin states to compensate flux dependent systematic uncertainties. Furthermore, the polarisation was destroyed and built up in reversed direction every four to five days. This compensates acceptance effects created by the dipole field necessary to sustain transverse target polarisation.

For the analysis events with incoming and outgoing muons and at least two reconstructed hadrons from the primary vertex inside the target cells are selected. Equal flux through the whole target is achieved by requiring that the extrapolated beam track crosses all cells. In order to select events in the DIS regime, cuts are applied on the squared four-momentum transfer, $Q^2 > 1\,(\mathrm{GeV}/c)^2$, and on the invariant mass of the final hadronic state, $W > 5\ \mathrm{GeV}/c^2$. Furthermore, the fractional energy transfer by the virtual photon is required to be $y > 0.1$ and $y < 0.9$ to remove events with poorly reconstructed virtual photon energy and events with large radiative corrections, respectively.

The hadron pair sample consists of all combinations of oppositely charged hadrons originating from the reaction vertex. Hadrons produced in the current fragmentation region are selected requiring $z > 0.1$ for the fractional energy of each hadron and $x_F > 0.1$. The Feynman variable $x_F$ is defined in the centre-of-mass frame of virtual photon and target nucleon as the longitudinal momentum of the hadron with respect to the virtual photon divided by the total available longitudinal momentum. The suppression of exclusive hadron pairs [27] is accomplished by requiring the missing mass to be $M_X > 2.4\,\mathrm{GeV}/c^2$. As the azimuthal angle $\phi_R$ is only defined for non-collinear vectors $\boldsymbol{R}$ and $\boldsymbol{q}$, a minimum value is required on the component of $\boldsymbol{R}$ perpendicular to $\boldsymbol{q}$, $|\boldsymbol{R}_\perp| > 0.07\,\mathrm{GeV}/c$. After all cuts, $5.8 \times 10^6$ $h^+h^-$ combinations for the $^6$LiD target and $10.9 \times 10^6$ $h^+h^-$ combinations for the $NH_3$ target remain. Figure 2 shows the invariant mass distributions of the two-hadron system for both targets, always assuming the pion mass for each hadron. A cut of $M_{hh} < 1.5\,\mathrm{GeV}/c^2$ is applied in order to justify the restriction to relative s and p waves given in Eqs. (8) and (9). For further details on the event selection and on the analysis we refer to Ref. [34].

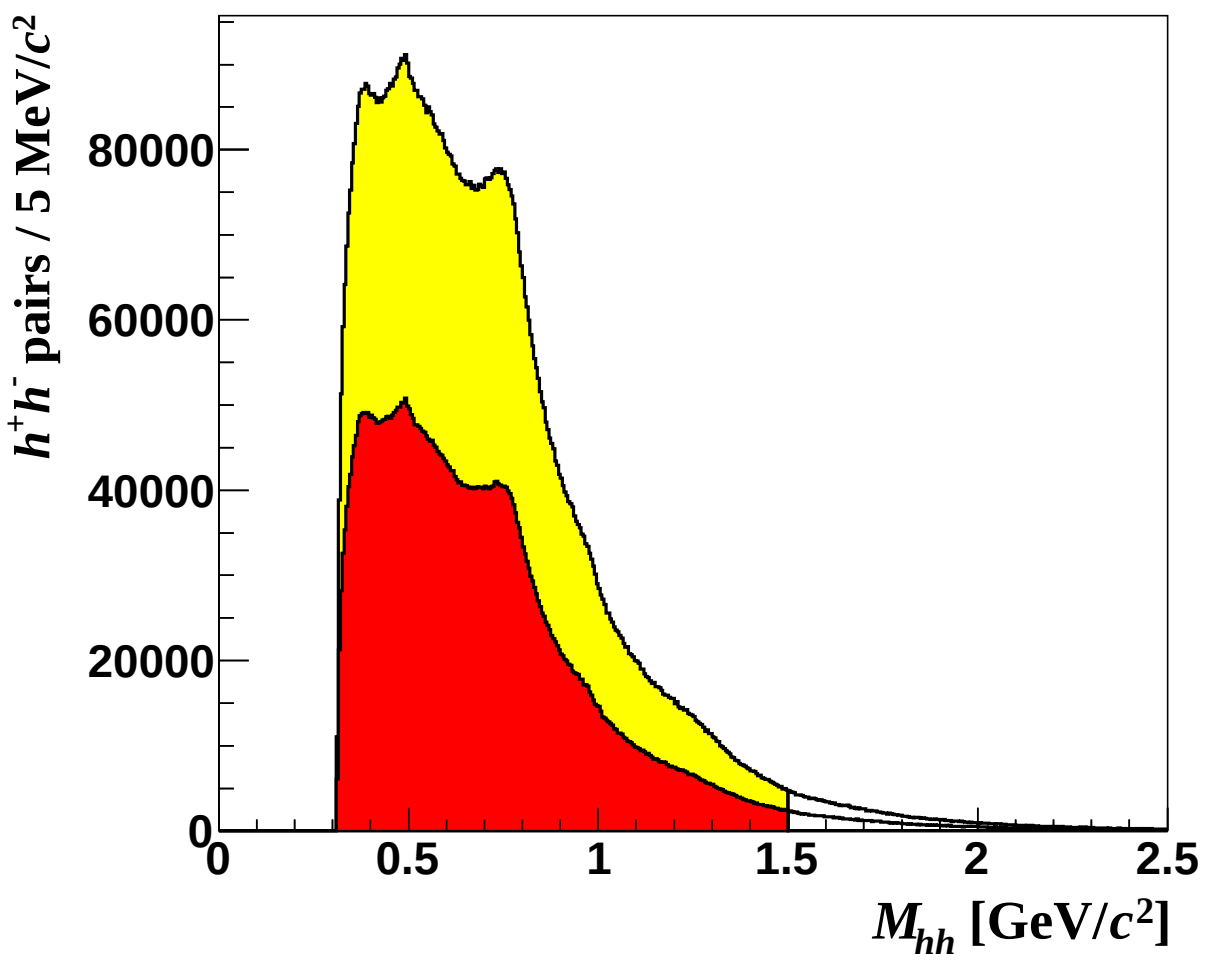

Fig. 2: Invariant mass distributions of the final $h^+h^-$ hadron samples for the $NH_3$ target (upper curve) and the $^6$LiD target (lower curve). The cut $M_{hh} < 1.5\,\mathrm{GeV}/c^2$ is indicated. In both cases the $K^0$, $\rho$ and $f_1$ resonances are clearly visible.

In Refs. [21, 22, 25], it was proposed to measure two-hadron production integrated over the angle $\theta$. This has the advantage that in the resulting expression for the cross sections only the fragmentation function $H_{1,ot}^{\sphericalangle}(z, M_{hh}^2)$ appears, provided that the experimental acceptance is symmetric in $\cos\theta$.

Figure 3 shows the $\sin\theta$, $\cos\theta$ and $\cos^2\theta$ distributions for the $NH_3$ target. The corresponding distributions for the $^6$LiD target show exactly the same shape. In the COMPASS acceptance, the opening angle $\theta$ peaks close to $\pi/2$ with $\langle\sin\theta\rangle = 0.94$ (Fig. 3, left). In the analysis, we extract the product $A = \langle A_{UT}^{\sin\phi_{RS}} \sin\theta\rangle$, integrated over the angle $\theta$. The $\cos\theta$ distribution is symmetric around zero (Fig. 3, centre) with $\langle\cos\theta\rangle = 0.01$. Therefore, the contribution to the asymmetry arising from $H_{1,lt}^{\sphericalangle}(z, M_{hh}^2)$ in Eq. (9) is expected to be rather small, so that the result is mainly sensitive to $H_{1,ot}^{\sphericalangle}(z,M_{hh}^2)$. The $\cos^2\theta$ distribution (Fig. 3, right) shows a mean value of $\langle\cos^2\theta\rangle = 0.11$. Therefore, the contribution of $D_{1,ll}(z,M_{hh}^2)$ in Eq. (8) does not vanish but contributes with a weight of about 16 % to the unpolarised cross section. This leads to a dilution of the asymmetry signal which has to be taken into account when extracting the transversity distributions from the data. For this purpose, the mean values of all three distributions for individual kinematic bins can be found on HEPDATA [35].

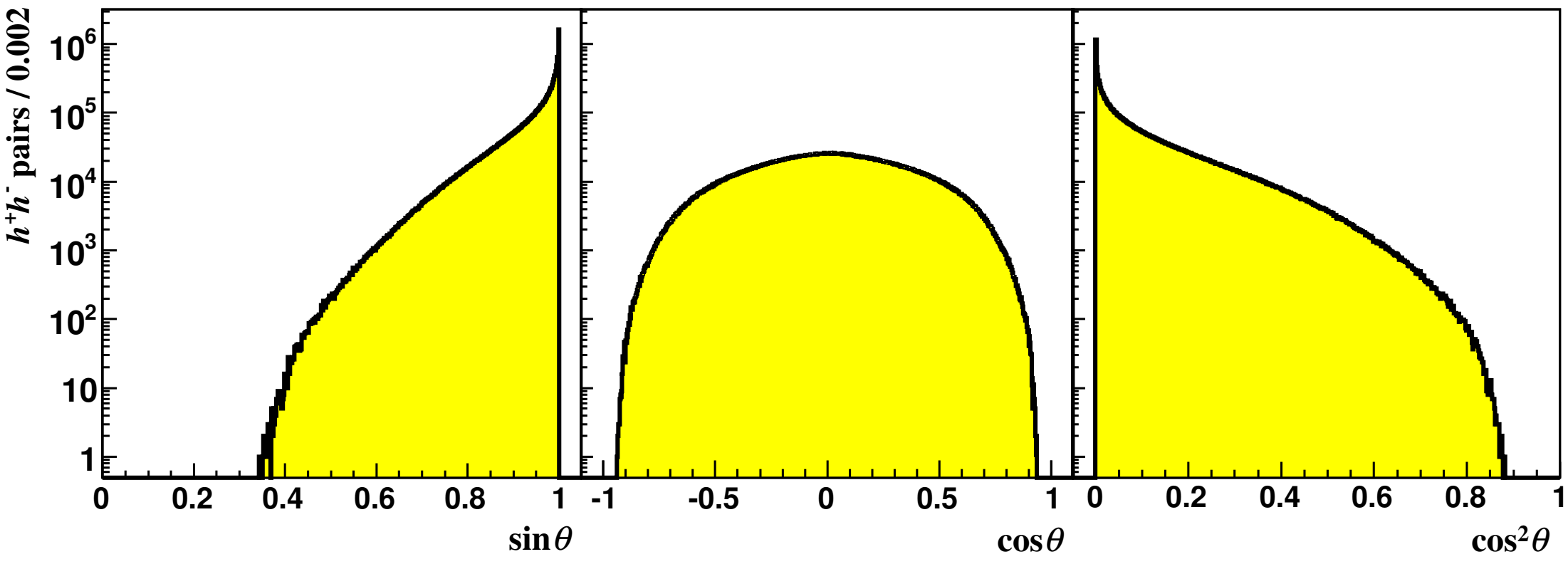

Fig. 3: Distributions in $\sin\theta$, $\cos\theta$ and $\cos^2\theta$ for the $h^+h^-$ sample from the transversely polarised $NH_3$ target.

## 4 Asymmetry Extraction

The asymmetry $\tilde{A} = \langle P_T\rangle A$ is evaluated in kinematic bins of $x$, $z$ or $M_{hh}$, while always integrating over the other two variables. As estimator an extended unbinned maximum likelihood function in $\phi_R$ and $\phi_S$ is used:

$$\mathscr{L} = \prod_{i=1}^{n_{cell}} \left\{ \left( e^{-I_i^+} \prod_{m=1}^{N_i^+} P^+(\phi_{Rm},\phi_{Sm};a_i^+,\tilde{A}) \right)^{\frac{\bar{N}}{N_i^+}} \left( e^{-I_i^-} \prod_{n=1}^{N_i^-} P^-(\phi_{Rn},\phi_{Sn};a_i^-,\tilde{A}) \right)^{\frac{\bar{N}}{N_i^-}} \right\}, \tag{10}$$

where the probability density function $P^\pm(\phi_R,\phi_S;a_i^\pm,\tilde{A}) = a_i^\pm(\phi_R,\phi_S)\cdot(1\pm f\cdot D_{nn}\cdot\tilde{A}\cdot\sin\phi_{RS})$ is normalised to the estimated number of hadron pairs $I_i^\pm = \int\int d\phi_R\, d\phi_S P^\pm(\phi_R,\phi_S;a_i^\pm,\tilde{A})$. Here $\pm$ denotes the sign of the target polarisation and $a_i^\pm(\phi_R,\phi_S)$ represents the acceptance seen by particles produced in target cell $i$, including the unpolarised cross section and the respective luminosities. The outer product corresponds to the target cells, $n_{cell} = 2$ in case of the $^6$LiD target and $n_{cell} = 4$ in case of the $NH_3$ target, while the inner products correspond to the two data samples of each cell $i$ acquired with target spin up and target spin down, respectively. The contributions are weighted with powers of $\bar{N}/N_i^\pm$ to account for unbalanced statistics, where $\bar{N}$ is the average number of pairs per sample and $N_i^\pm$ is the number of pairs with spin up or spin down for each target cell $i$. The weighting makes the resulting asymmetries less sensitive to acceptance and reduces possible false asymmetries. Monte Carlo studies have shown that the functional form of the acceptance has a negligible effect on the extracted asymmetries. Hence constant values $a_i^\pm$ are used, which account for the different luminosities in the two periods with opposite target polarisation. Additionally, each pair is weighted by the corresponding target dilution factor $f(x,y)$ and the transverse spin-transfer-coefficient $D_{nn}(y)$. The magnitude of the target polarisation cannot be used as a weight in the fit since this could bias the results in case of unbalanced statistics, instead its mean value is used to scale the extracted asymmetries $\tilde{A}$. In order to avoid false asymmetries, care was taken to select only such data for the analysis for which the spectrometer performance was stable in consecutive periods of data taking. This was ensured by extensive data quality tests. In a first step, the detector performance was investigated on the time scale of a SPS extraction, typically 4.8 s every 16.8 s. Quantities directly linked to detection and reconstruction stability were studied, like number of interaction vertices, number of tracks, number of clusters in the calorimeters and trigger rates. Time intervals with irregularities in these variables were discarded from further analysis. In a second step, the distributions of DIS and SIDIS variables were investigated. In order to ensure sufficient statistics, this was done on the basis of a run that typically consists of 200 beam extractions. Runs showing distributions that are statistically incompatible to the majority of runs were also discarded from the analysis. The remaining data sample was carefully scrutinised for a possible systematic bias in the final asymmetries. Here, the two main sources for uncertainties are false asymmetries, which can be evaluated by combining data samples with same target spin orientation, and effects of the acceptance, which can be evaluated by comparing subsamples corresponding to different ranges in the azimuthal angle of the scattered muon. No significant systematic bias could be found. Therefore, an upper limit was estimated comparing the results of the systematic studies to expected statistical fluctuations. The resulting systematic uncertainty for each data point amounts to about 75 % of the statistical error for both targets. An additional scale uncertainty of 5.4 % for the $^6$LiD and 2.2 % for the $NH_3$ target accounts for uncertainties in the determination of target polarisation and target dilution factor calculated for semi-inclusive reactions [36].

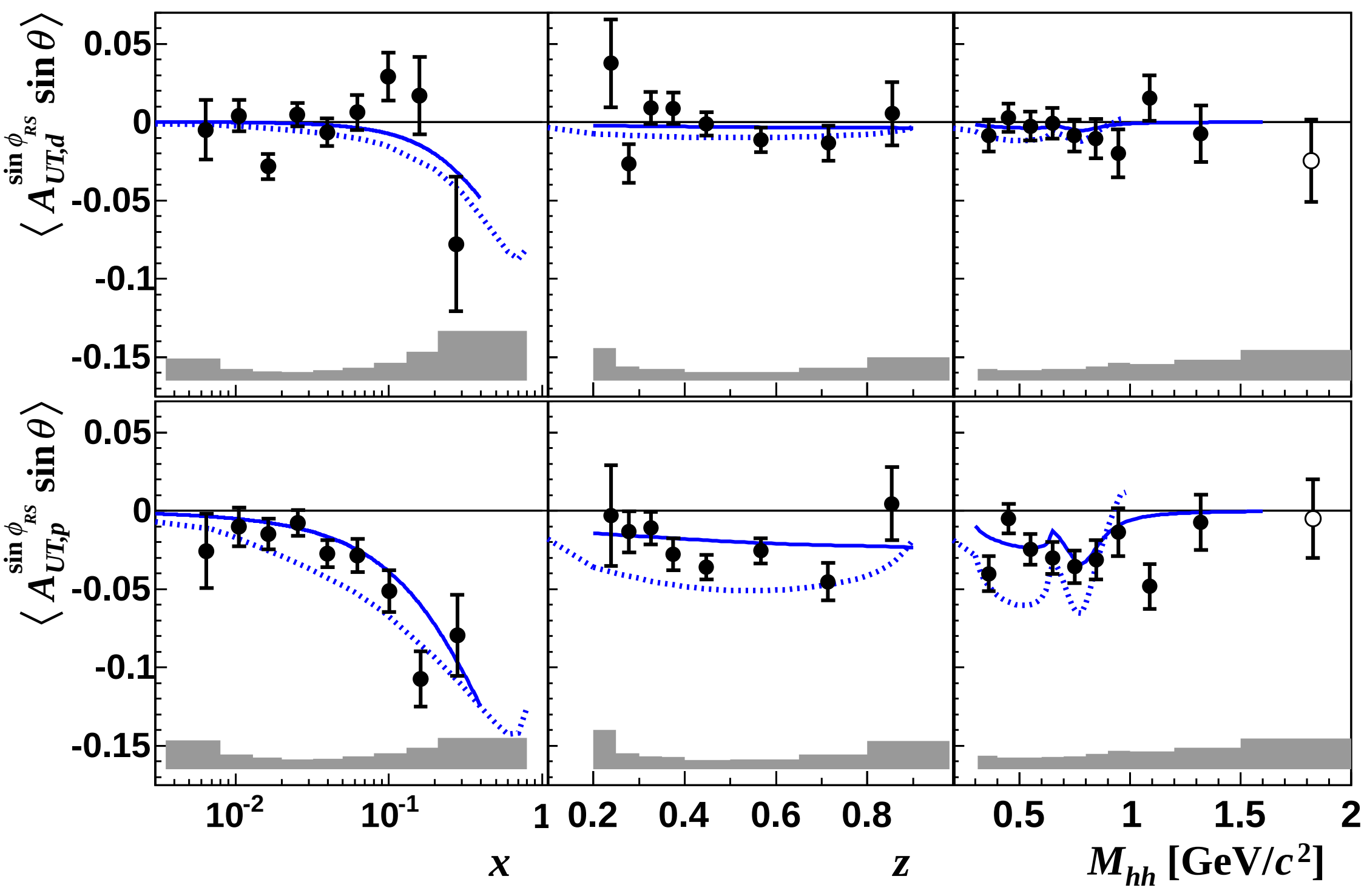

Fig. 4: Deuteron and proton asymmetries, integrated over the angle $\theta$, as a function of $x$, $z$ and $M_{hh}$, for the data taken with the $^6$LiD (top) and $NH_3$ target (bottom), respectively. The open data points in both asymmetry distributions vs. $M_{hh}$ include all hadron pairs with an invariant mass of $M_{hh} \geq 1.5$ GeV/$c^2$. These pairs are discarded for the two other distributions, which are integrated over $M_{hh}$. The grey bands indicate the systematic uncertainties, where the last bin in $M_{hh}$ is not fully shown. The curves show the comparison of the extracted asymmetries to predictions [37, 38] made using the transversity functions extracted in Ref. [15] (solid lines) or a pQCD based counting rule analysis (dotted lines).[1]

## 5 Discussion of Results

The resulting asymmetries are shown in Fig. 4 as a function of $x$, $z$ and $M_{hh}$ for the $^6$LiD (top) and $NH_3$ (bottom) targets, respectively. For $^6$LiD, no significant asymmetry is observed in any variable. For $NH_3$, large negative asymmetries are observed in the region $x > 0.03$, which implies that both transversity distributions and polarised two-hadron interference fragmentation functions do not vanish. For $x < 0.03$, the asymmetries are compatible with zero. Over the measured range of the invariant mass $M_{hh}$ and $z$, the asymmetry is negative and shows no strong dependence on these variables.

When comparing the results on the $NH_3$ target to the published HERMES results on a transversely polarised proton target [28], the larger kinematic region in $x$ and $M_{hh}$ is evident. However, both results cannot be directly compared for several reasons: (1) The opposite sign is due to the fact that in the extraction of the asymmetries the phase $\pi$ in the angle $\phi_{RS}$ is used in the COMPASS analysis; (2) COMPASS calculates asymmetries in the photon-nucleon system, while HERMES published them in the lepton-nucleon system; both agree reasonably well when including $D_{nn}$ corrections for HERMES; (3) HERMES uses identified $\pi^+\pi^-$ pairs and COMPASS $h^+h^-$ pairs; (4) COMPASS applies a minimum cut on $z$, removing a possible dilution due to contributions from target fragmentation.
A naive interpretation of our data, based on Eq. (7) and on isospin symmetry and charge conjugation, yields $D_{1,u} = D_{1,d}$ and $H^{\sphericalangle}_{1,u} = -H^{\sphericalangle}_{1,d}$ [27]. When considering only valence quarks, the asymmetry $A^{\sin\phi_{RS}}_{UT,d}$ is proportional to $[h^u_1 + h^d_1]H^{\sphericalangle}_{1,u}$ for the deuteron target, while for the proton target $A^{\sin\phi_{RS}}_{UT,p} \propto [4h^u_1 - h^d_1]H^{\sphericalangle}_{1,u}$. Therefore, like in the case of the Collins asymmetry, the small asymmetries observed for the deuteron target imply $h^u_1 \approx -h^d_1$, while the sizeable asymmetries for the proton target imply a non-vanishing $u$ quark transversity $h^u_1$.
In an early theoretical approach, a strong s and p wave interference in two-pion production, known from two-meson phase shift analyses, was considered for a transversely polarised proton target [21]. As a result, a sign change of the two-pion asymmetries in the vicinity of the $\rho$ mass was predicted. From our data, however, such a sign change can clearly be excluded.
In a different approach [25–27], all two-hadron fragmentation functions for two-pion production were calculated in the framework of a spectator model for the fragmentation process. In the mass range up to $M_{hh} = 1.3$ GeV/$c^2$, all relevant two-pion channels with relative s and p waves were considered and the parameters of the model were tuned to fit the output of the PYTHIA [39] event generator tuned to HERMES kinematics. Predictions were made for the IFF $H^{\sphericalangle}_1$ as well as for $D_{1,ol}$ and $D_{1,ll}$ and in Ref. [27] the expected asymmetries for COMPASS using deuteron and proton targets were calculated assuming different models for the transversity distributions. Very recently [37], these parametrisations of the two-hadron fragmentation functions were used together with the transversity distributions extracted from single-hadron production [15] to make predictions for both proton and deuteron targets using the kinematic range covered by COMPASS. The applied cuts of $0.2 < z < 0.9$, $0.3 < M_{hh} < 1.6$ GeV/$c^2$, $0.003 < x < 0.4$, and $0.1 < y < 0.9$ assure a wide overlap with the presented data. The solid lines in Fig. 4 show the good agreement of these predictions in comparison to the data obtained using the $^6$LiD and $NH_3$ targets. For $^6$LiD, the predicted asymmetries vs. $z$ and $M_{hh}$ are smaller than 0.01, in good agreement with the data. Only for $x > 0.1$ the predicted asymmetries start to exceed 0.01, a trend which can neither be confirmed nor excluded by the data. For $NH_3$, significant asymmetries are predicted and especially the shape depending on $x$ is well described. Also the dependences on $z$ and $M_{hh}$ are reasonably described, showing a rather weak dependence on $z$ as well as the enhanced asymmetries close to the $\rho$ mass. The predicted kink at the $\rho$ mass is not visible in the data but can also not be excluded.
The same parameterisations of the two-hadron FF were also used in Ref. [38], together with two models for the transversity distributions, a SU(6) quark-diquark model and a pQCD-based counting rule analysis. For the latter, the results using proton or deuteron targets are shown by the dotted lines in Fig. 4. The cuts applied in Ref. [38] were similar to those used in Ref. [37], except that the invariant mass range was restricted to $M_{hh} < 1.0$ GeV/$c^2$, the $z$ region was extended to $0.1 < z < 0.9$ and no explicit $x$ cut was applied. The resulting predictions describe the data fairly well. While for the $^6$LiD target the predicted asymmetries are still in agreement with the data, for the $NH_3$ target they tend to slightly overestimate the measured asymmetries.

[1] The sign of the original predictions was changed to accommodate the phase $\pi$ in the definition of the angle $\phi_{RS}$ used in the COMPASS analysis.

## 6 Conclusions

In this Letter we present the results on azimuthal asymmetries measured in two-hadron production in semi-inclusive deep inelastic scattering using transversely polarised $^6$LiD (deuteron) and $NH_3$ (proton) targets. These asymmetries provide independent access to the transversity distribution functions complementary to that using the Collins effect in single-hadron production. For the deuteron target, no significant asymmetries are observed. For the proton target, sizeable asymmetries are measured in all three kinematic variables, $x, z$ and $M_{hh}$, indicating non-vanishing $u$-quark transversity and two-hadron interference fragmentation functions. No dependence of the asymmetries is observed for the variables $z$ and $M_{hh}$, on which the interference fragmentation function $H^{\sphericalangle}_1$ directly depends. Especially, the change of sign in the vicinity of the $\rho$ mass predicted by Ref. [21] is excluded. The dependence of the asymmetry on the Bjorken variable $x$, observed on the proton target, constitutes independent experimental informa-tion on the $u$-quark transversity distribution function $h_1^u(x)$. In conjunction with the unique results from the deuteron target, also the $d$-quark transversity function $h_1^d(x)$ can be accessed. In an interpretation based on valence quarks only, the proton data indicate the $u$-quark transversity function to be sizeable, while the deuteron data indicate an approximate cancellation of the $u$- and $d$-quark transversity functions.

We acknowledge the support of the CERN management and staff, as well as the skills and efforts of the technicians of the collaborating institutes. Further we want to thank A. Bacchetta and M. Radici as well as M.-Q. Ma for kindly providing predictions of the asymmetries adapted for COMPASS kinematics.